\documentclass[conference,compsoc]{IEEEtran}

\usepackage{cite}

\ifCLASSINFOpdf
\else
\fi
\usepackage{algorithmic}

\usepackage{array}
\usepackage{graphics}
\usepackage{multirow}
\usepackage{epsfig}
\usepackage{subfigure}
\usepackage{amsmath}

\begin{document}

%
\title{Fully Dense Neural Network for the Automatic Modulation Recognition}

\author{Miao~Du,
Qin~Yu,
Shaomin~Fei,
Chen~Wang,
Xiaofeng~Gong,
and~Ruisen~Luo*
\IEEEcompsocitemizethanks{
\IEEEcompsocthanksitem Miao Du, Qin Yu, Chen Wang, Xiaofeng Gong, Ruisen Luo are with College of Electrical Engineering, Sichuan University, 24 South Section 1, One Ring Road, Chengdu, China, 610065. Chen Wang is now based on Department of Computer Science, Rutgers University -- New Brunswick, Piscataway, New Jersey 08854, USA; the work was done when Chen Wang was at Sichuan University. Shaomin Fei is with Engineering Practice  Center,  Chengdu  University  of  Information Technology \protect\\
\IEEEcompsocthanksitem * the Corresponding Author, Ruisen~Luo, 
Email: rsluo@scu.edu.cn.}
}
\IEEEtitleabstractindextext{%
\begin{abstract}
Nowadays, we mainly use various convolution neural network (CNN) structures to extract features from radio data or spectrogram in AMR. Based on expert experience and spectrograms, they not only increase the difficulty of preprocessing, but also consume a lot of memory. In order to directly use in-phase and quadrature (IQ) data obtained by the receiver and enhance the efficiency of network extraction features to improve the recognition rate of modulation mode, this paper proposes a new network structure called Fully Dense Neural Network (FDNN). This network uses residual blocks to extract features, dense connect to reduce model size, and adds attentions mechanism to recalibrate. Experiments on RML2016.10a show that this network has a higher recognition rate and lower model complexity. And it shows that the FDNN model with dense connections can not only extract features effectively but also greatly reduce model parameters, which  also provides a significant contribution for the application of deep learning to the intelligent radio system.
\end{abstract}

\begin{IEEEkeywords}
Feature multiplexing, Signal Attention; Automatic modulation recognition.
\end{IEEEkeywords}
}

\maketitle

\IEEEdisplaynontitleabstractindextext

%

%
\IEEEpeerreviewmaketitle

\section{Introduction}

\IEEEPARstart{I}{n} communication system, the data must be modulated using carriers in a specific frequency range to make it transmit information efficiently over long distances. In receiving end, automatic modulation recognition (AMR) play a important role in many civil and military fields. In military applications, radio signals received by the air environment in non-cooperative situations need recognition the types of modulation. In civil applications, multiple modulation types are implemented with a signal transmitter to control data rate and line security. If a wireless system can automatically identify modulation types, it can save communication protocol header resources.

There are usually two methods for modulation recognition of radio: likelihood-based methods \cite{sills1999maximum_1} and the hand-craft feature extraction with expert experience  \cite{nandi1998algorithms_2} \cite{dai2002joint_3}. Previous works comparing these two approach have found that feature-based approach could achieve good performance with lower computational cost. In the early stage of feature extraction, statistical features were used for AMR, such as high-order statistics \cite{swami2000hierarchical_4} \cite{xie2019deep_5} and periodic stationary features \cite{hong2005classification_6}, combined with different algorithms for pattern recognition, such as decision trees , support vector machines, and artificial neural networks. Many extraction methods and classification algorithms are described and applied to various modulation signals \cite{zhu2015automatic_7}. By automatically classifying the modulation type on the receiver side, AMR has been part of intelligent radio systems.

Recently, deep learning has shown very good performance in the field of image and speech, and has achieved successful applications. In order to build a more intelligent and flexible communication system, a few literatures have studied the application of deep learning in modulation recognition. GNU was used to generate IQ data containing 11 different modulated signals, where the IQ data was recognized by the convolutional neural network (CNN) for the first time \cite{o2016convolutional_8}. The importance of the data set was discussed in \cite{o2016radio_9}. And the RML2016.10a data set was proposed and a baseline of the CNN2 on this data set was given. Later, \cite{west2017deep_10} proposed CLDNN, and discussed the effect of ResNet under various hyperparameters in the RML2016.10a dataset, and proved that the recognition rate was not limited by the network depth, but they were limited by features purely CNN architectures can learn. Recently \cite{zhang2018automatic_11} combined the raw IQ data and Fourth order Cumulants (FOC) together to represent the modulated signal, and proposed CNNR-IQFOC structure to improve recognition rate. On other method, \cite{jeong2018spectrogram_12} explores the IQ data into a spectrum in simulation experiments. \cite{zeng2019spectrum_13} \cite{daldal2019automatic_14} transformed signals into spectrogram images using the short-time discrete Fourier transform, and save spectrogram images for recognition. It is worth mentioning that \cite{west2017deep_10} \cite{luo2019radio_15} studied the combination of CNN and Long Short-Term Memory (LSTM) cell to extract features in the RML2016.10a. \cite{west2017deep_10} used skip connection to reuse the features extracted from the previous CNN layer, concatenating them into the latter layer, and finally input the multi-layer information into LSTM. \cite{zheng2019fusion_16} used ResNet and passed features to the next layer. While these methods got good results, they did not discuss and discover that the feature map dense connection is promising in AMR. 

In general, there are three methods to use IQ data which received by in AMR: 1. Use IQ data directly ; 2. Use some expert knowledge; 3. Convert data to spectrograms or constellation diagrams \cite{wang2019data_17} \cite{peng2018modulation_18}. However, extracting features requires a strong expert knowledge, and using spectrograms or constellation diagrams will increase the complexity of data processing and consume more computer memory. In order to try not to use the preprocessing and improve the efficiency of network extraction features, this paper proposes a new network structure called Fully Dense Neural Network (FDNN). 

\begin{figure*}[htbp]
\centering
\includegraphics{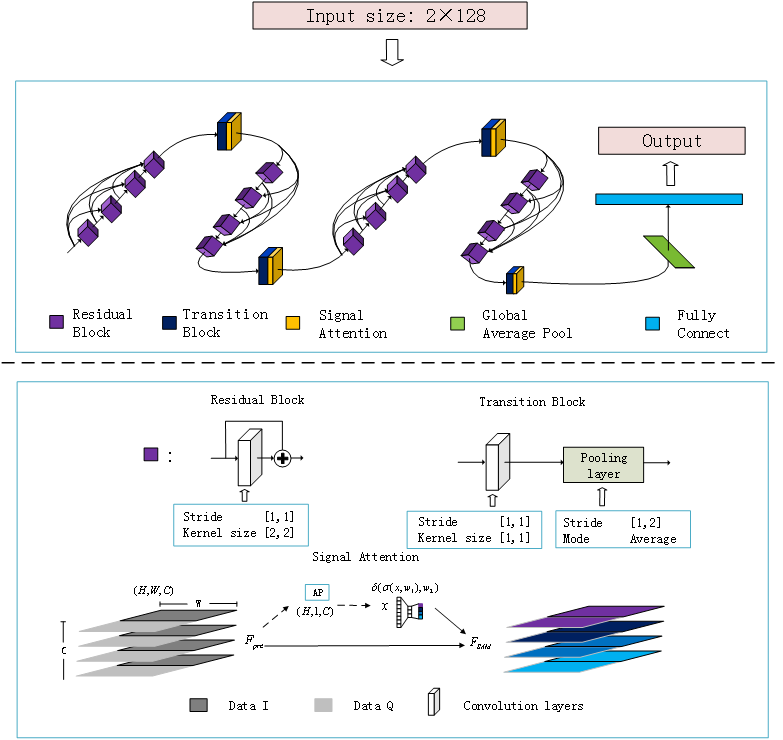}
\caption{The structure of Fully Dense Neural Network.}
\label{fig_sim2}
\end{figure*}

Since the data can be extracted different features with convolution kernels while training the neural network, the contribution of each feature maps are different. So in order to make the network focus on more important feature and recalibration the network, we improved a Squeeze-and-Excitation Blocks as Signal Attention \cite{hu2018squeeze_19}. And in the network residual blocks used for getting features. This is known as a residual network (ResNet) because the forwarded information forces the network to learn a residual function as part of feature extraction \cite{he2016deep_20}. In addition to the residual block, the dense structure not only extracts features more effectively, but also reduces the complexity of the network while encouraging feature reuse to improve training efficiency \cite{huang2017densely_21}. The structure of Fully Dense Neural Network shows in Fig. \ref{fig_sim2}

Through a lot of experiments with RML2016.10a, we not only prove that the use of our network can greatly improve the recognition performance , but also shows that the proposed network is simpler, more powerful and superior performance.

\section{Fully Dense Neural Network}

\subsection{Modulated Signal Model} 

\begin{figure}[h]
\centering
\includegraphics[scale=0.5]{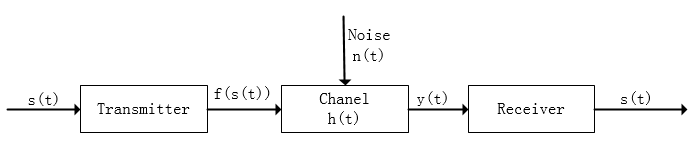}
\caption{the simple communication process of the radio system.}
\label{fig_sim1}
\end{figure}

We can see the simple communication process of the radio system from Fig. \ref{fig_sim1}, including: the transmitting end, the transmitting channel, the noise channel and the receiving end. y(t) is recorded as the last received data, expressed as:
\begin{equation}
y(t)=f(s(t)) * h(t)+n(t)
\end{equation}
where s(t) is the source signal, t is the time size, h is the channel response, and n is the noise. It is one of the tasks of this paper to use depth learning to identify the modulation of the received signal y(t). Common 11 modulation methods \cite{zhu2015automatic_7}, such as: BPSK, QPSK, 8PSK, 16QAM, 64QAM, BFSK, CPFSK, PAM4, WB-FM, AM-SSB, and AM-DSB.

\subsection{Methods} 
\subsubsection{Signal Attention Mechanism}

Attention can be viewed, broadly, as a tool to bias the allocation of available processing resources towards the most informative components of an input signal. In order to make the network focus on the more important part in AMR, this paper uses Signal Attention Mechanism as shown in Fig. \ref{fig_sim2}. We can see that $F_{p r e} \longrightarrow F_{S A M}$, $F_{p r e} \in R^{H \times W \times C}$, $F_{SAM} \in R^{H^{\prime} \times W^{\prime} \times C}$, where $F_{pre}$ is the output of the previous layer, $F_{SAM}$ is the result of using signal attention mechanism.

In order to retain I and Q information, we used an average pooling with $W$ strides, compressing each row to a value, and then using the fully connected layer to get the weight of each channel. Finally the signal attention:

\begin{equation}
x=\left(\frac{1}{W} \sum_{i=1}^{W} F_{p r e}^{i}(0, i), \frac{1}{W} \sum_{i=1}^{W} F_{p r e}^{i}(1, i)\right)
\end{equation}

\begin{equation}
F_{S A M}=\delta\left(\sigma\left(x, w_{1}\right), w_{2}\right)
\end{equation}

Where $\boldsymbol{F}^{i}_{\text{pre}}$ is $i-\text{th}$ channel of the previous layer $\boldsymbol{F}_{\text{pre}}$. For learning the nonlinear relationship between channels, \cite{hu2018squeeze_19} designed a simple gating mechanism with a sigmoid activation, where $\sigma$ refer to the Relu \cite{nair2010rectified_22} function, $w_{1} \in R^{2 c \times \frac{2 c}{r}}$, $w_{2} \in R^{\frac{2 c}{r} \times c}$. There has a reduction ratio $r$ in $w_{1}$, $w_{2}$, which encode-decode the $w$ and reduces the trainable parameters.

\subsubsection{Dense Structure and Residual Block Theory}

Convolution neural networks are a commonly used neural network for extracting features. In a standard feed forward network, the output of the $n \operatorname{th}$ layer is noted as $\mathrm{x_{n}}=\mathrm{H}(\mathrm{x_{n-1}})$, where $x_{n-1}$ is the input of $n \operatorname{th}$ layer and the output $x_{n}$ of the $n \operatorname{th}$ layer is obtained by nonlinear transformation $\mathrm{H}$. In Fig. \ref{fig_sim2}, in order to make the trained network deeper, the residual block proposed a skip connection that adds the input to the nonlinear transformation:

\begin{equation}
\mathrm{x_{n}}=\mathrm{H}(\mathrm{x_{n-1}})+\mathrm{x_{n-1}}
\end{equation}

This jumped connection allows the gradient to propagate directly to the previous layer, making training easier. Densenet [12] continues to enhance this jump connection, concatenate the previous layer output to all subsequent layers:

\begin{equation}
x_{l}=H_{l}\left(\left[x_{0}, x_{1}, \ldots, x_{l-1}\right]\right)
\end{equation}

where $\left[x_{0}, x_{1}, \ldots, x_{l-1}\right]$ is made up of the characteristics of the front layer. This dense connection not only allows the gradient energy to propagate to the previous layer, but also allows the features to be reused in the next layer, improving the efficiency of the feature. 

In the FDNN, there are three main modules: the Residual Block, the Transition Block, and the Signal Attention Block. In the Residual Block, we use the classical residuals structure which include 'Relu' functions, Batch Normazation (BN) \cite{ioffe2015batch_23}, and convolutional layer. In the Transition Block, Batch Normazation, 'Relu', convolutional layer and average pooling layer are included. The network is designed with the growth parameter $K$ and reduction rate $\theta$. In the function $H_{l}$, it produces $K$ feature maps and follows the $l$th layer has $k_{0}+k\times(l-1)$, where $k_{0}$ is the number of channels in the input layer. However, as the number of FDNN layers increases, feature maps in per layer becomes larger and larger. So the Transition Block also is used to control the number of the convolutional feature maps by parameters $\theta$. 

\begin{table}[t]
\centering
\caption{The parameters for the weights of the FDNN model.}
\begin{tabular}{l l}
\hline

Parameters  & values                 \\ \hline
Max Epochs & 40\\
Mini Batch Size & 64\\
Initial Learn Rate  & 0.001\\
Early Stop    &    10\\
Learn Rate Drop Factor   &    0.4\\
Learn Rate Drop Period     &   5\\
In FDNN: \\
~ ~ Growth Parameter  &  15\\
~ ~ Reduce Rate  & 0.5\\\hline

\end{tabular}
\end{table}

When using IQ data for modulation recognition, the input is $x \in R^{2 \times n}$ , where 2 consists of a row I and a row Q, and n represents the length of the received signal, so if the input x is treated as an image, the value of this image pixel is represented amplitude, the length of the image represents the time series. In this paper, we consider that I and Q are not independent, the $(2,2)$ convolution kernel is used when extracting features, but the transition block uses a one-dimensional pooling layer to preserve signal information. To enhance feature reuse, we introduce dense connections to the network. However, as can be seen from the extensive AMR literature, the use of residuals is beneficial for feature extraction. In block-to-block feature transfer, in order to learn the weight of transferred feature maps, we also added the full connection process of encoding-decoding, which is conducive to network reconstruction. In this network, there are not only intensive multiplexing features at each layer, but also simple addition of features within the layer. So this network is called as Fully Dense Neural Network.

\section{Experiments} 

\subsection{Data}
RadioML2016.10a standard radio signal data set is used for training and testing data. The data set includes 11 different types of modulation signal data. It considers 11 modulation methods: BPSK, QPSK, 8PSK, 16QAM, 64QAM, BFSK, CPFSK, PAM4, WB-FM, AM-SSB and AM-DSB, which are widely used in practical communication systems. Each data frame is $2 * 128$ in size, totaling about 220,000 frames.

\subsection{Experiment}

To evaluate the recognition performance of the proposed network in IQ data, we compare the recognition accuracy of the FDNN with four methods from \cite{o2016convolutional_8}, \cite{west2017deep_15}, \cite{zhang2018automatic_11} and \cite{west2017deep_10}, which we refer to as CNN2-IQ, Resnet, CNNR-IQFOC and CLDNN, respectively. We implemented neural networks in the , where FDNN-IQ is the FDNN without the Signal Attention Mechanism and the Densenet has 32 layers.

This paper uses keras to design FDNN with 32 layers for AMR, while the ResNet has 34 layers noted as Resnet-34. In the experiment, the GTX 1080Ti is used to train and test these networks. In the preparation of data, we randomly selected 200 signals per modulation mode per SNR as the test set and the remaining signals as the validation data (100 signals per modulation mode per SNR) and the training data (700 signals per modulation mode per SNR). During the training, The learning rate was 0.001, K was 15, the batch size was 64, the epochs are 40, and early stop was 10. In other words, if the recognition rate of the verification set is not improved in 10 cycles, the training will be stopped. More details in Table 1. Since we did not adjust these parameters in the experiment, it might have been better performance to fine-tune them.

\begin{figure}[htbp]
\centering
\includegraphics[width=9cm,height=6cm]{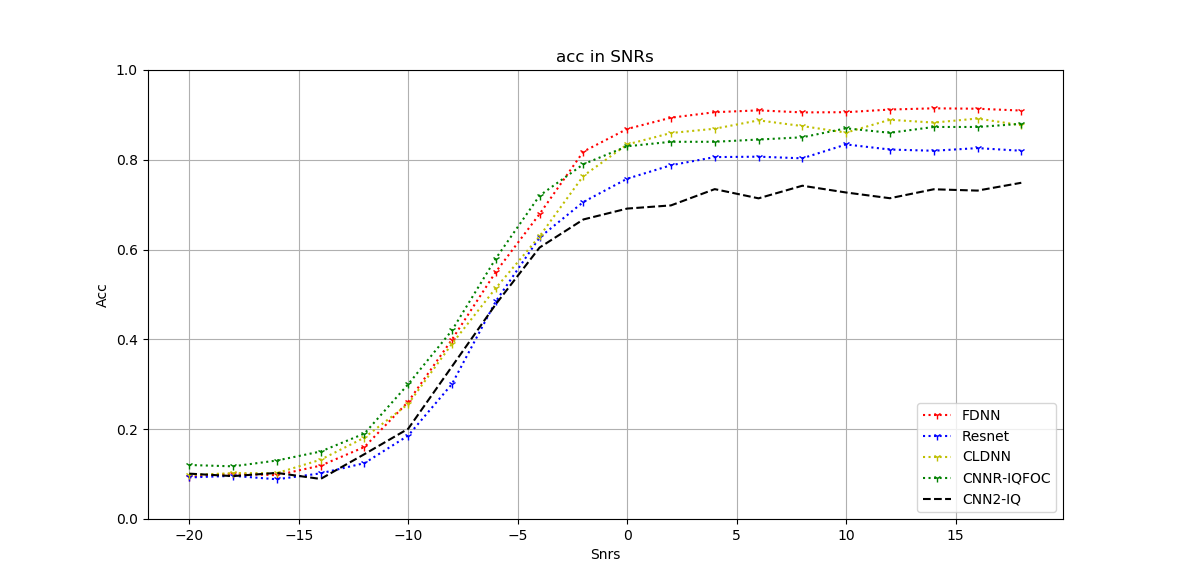}
\caption{the recognition performance of CADRN, CNN2-IQ and
CNNR-IQFOC under various signal-to-noise ratios (Snrs). The x-axis represents the Snrs and the y-axis represents the accuracy.}
\label{fig_sim3}
\end{figure}

\begin{table*}[htbp]
\centering
\caption{shows the average recognition of each model at 0~18dB . ( Resnet-50 : standard ResNet-50 \cite{he2016deep_20} ; FDNN :Fully Dense Neural Network ( FDNN) ;  FDNN-IQ: FDNN without attentions; Paras : Trainable Parameters; Time1:  The consumed time for training of the network represents only one training epoch; Time2:  The time for testing of the network represents only one signal )}
\begin{tabular}{|c|c|c|c|c|c|c|c|c|}
\hline

Methods  & CNN2    & Resnet-34 & Resnet-50 & CNNR- IQFOC & CLDNN & Densenet & DNN-IQ & FDNN                 \\ \hline
Accuracy(\%) & 72.4  & 80.8 & 84.4   & 85.6 & 87.2 & 88.3 & 88.5 & 89.6        \\ \hline
Paras(K)    & 2,308  & 204 & 23,550   & 3,675 & 181 & 109 & 40 & 45.5        \\ \hline
Time1(mins)   & 0.82   & 2.28 & 5.5   & - & 4.77 & 3.90 & 3.68 & 5.5        \\ \hline
Time2(ms)   & 0.15   & 0.20 & 0.76   & - & 1.07 & 0.37 & 0.47 & 0.49        \\ \hline

\end{tabular}
\end{table*}

\section{Results and discussions} 

\subsection{Results}

Fig. \ref{fig_sim3} shows the comparison of recognition rates of models at various SNR. The recognition rate of Resnet-34 and CNN2-IQ is significantly lower than that of FDNN. In Fig. \ref{fig_sim3} shows that the recognition accuracy of the FDNN is around higher 3\%-6\% than those of the CNNR-IQFOC when SNR is above 0 dB. At 6 dB, the FDNN gets around 6.6\% higher accuracy than CNNR-IQFOC. At 18 dB, the FDNN gets around 2.5\% higher accuracy than CNNR-IQFOC. FDNN has no obvious advantage over CLDNN at low SNR, and starts to surpass CLDNN in recognition rate when SNR is -4. However, when the SNR is below 0 dB, the FDNN recognition rate is about 2\%-3\% lower than CNNR-IQFOC. It is noted that the FDNN has better recognition performance at high SNR compared with ResNet-34, CNN2-IQ and CLDNN. It suggests that the FDNN is more effective than other networks in extracting features at high SNR.

Table. 2 shows the average recognition rate from 0 to 18 dB is 85.6\% with CNNR-IQFOC, while the average recognition rate with FDNN is 89.6\%. If the Signal Attention Mechanism is not used in FDNN, the recognition rate drops by 1.1\%. Compared with CNN2-IQ, our proposed network has significant advantages. Meanwhile, we can find that the trainable parameters in FDNN are 46 K, while the trainable parameters in CNNR-IQFOC are 3,675 K. We present the model with fewer trainable parameters than CNNR-IQFOC about 80 times, while the average recognition rate is higher than 4\%. Among them, the standard 50-layer residual network (Resnet-50) was added for the experiment, and the result showed that the recognition rate of the Resnet-50 was 5.2\% lower than FDNN , and the trainable parameters increased by about 61.5 times. Although Resnet-34 has a smaller trainable parameter than Resnet-50, the recognition rate is 8.8\% lower than our proposed model. The experiment proved that the recognition rate is not limited by the network depth, and residual block is an effective component in learning features. On other hand, our network structure spent test time was moderate, while  required more time per epoch during training. Compared with the CNN structure, although there is a certain improvement in recognition rate when LSTM is used to train the network, it takes more time to train and test.

In the experiment, Densenet is a more effective network than Resnet. Compared with those networks without skip connect, feature reuse can greatly reduce the trainable parameters of the network and the recognition rate is improved by fully used features. The experiment suggests that the FDNN is a network structure that can effectively extract features.

\begin{figure}[tbp]
\centering
\includegraphics[width=9cm,height=6cm]{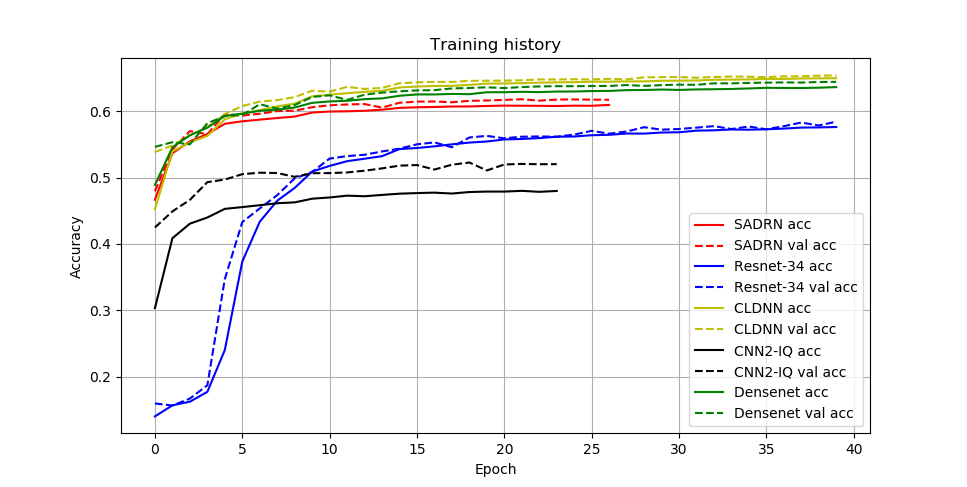}

\caption{Training history showing the the training accuracy and validation accuracy.}
\label{fig_sim4}
\end{figure}

In Fig. \ref{fig_sim4} shows the training history about the training accuracy and validation accuracy. CNN2-IQ and FDNN appeared early stop at 23 and 26 epoch respectively, while other models are trained to 40 epoch. In recognition rate and loss value of verification set, Densenet and CLDNN surpass FDNN. However, the recognition rate of the test set is lower than that of FDNN, which indicates that the Densenet and CLDNN have a situation of overfitting.

\begin{figure*}[h]
\centering
\subfigure[SA1: the first block of attention ]{\includegraphics[width=3in]{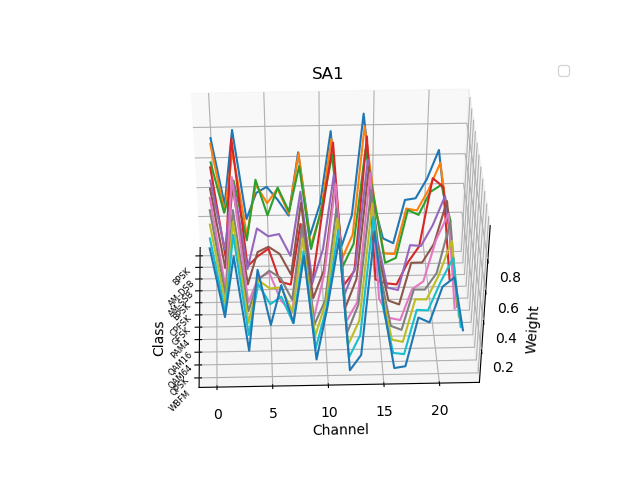}}
\subfigure[SA4: The fourth block of attention]{\includegraphics[width=3in]{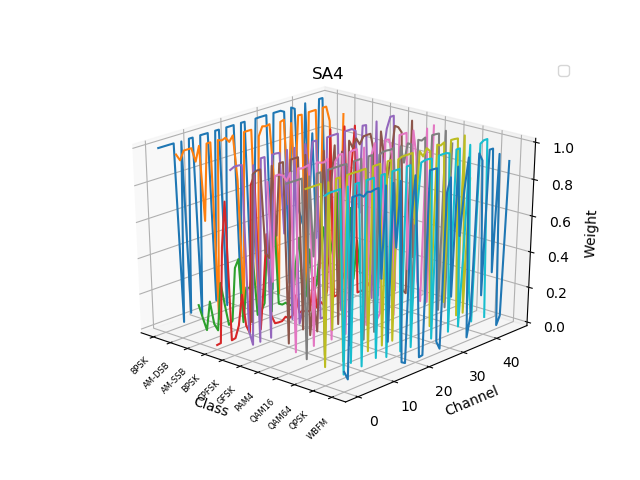}}
\caption{These 3D maps represent changes in attention in the neural network. SAn presents the nth of Signal Attention (SA) , where the larger n means the higher layers. The three dimensions are respectively represented as category, channel and weight.}
\label{fig_sim5}
\end{figure*}

\subsection{Discussion about Signal Attention}

To understand the role of the Signal Attention (SA) in the network, we selected 11 kinds of modulation signals with SNR of 18 for visualization, as shown in Fig. \ref{fig_sim5}. We can see that there are different weights for different feature maps in our network in Fig. \ref{fig_sim5}. In lower layers, most categories of attention are similar, such as there are five more important feature maps which are $0 \mathrm{th}$, $2 \mathrm{th}$, $8 \mathrm{th}$, $11 \mathrm{th}$,$14 \mathrm{th}$,$21 \mathrm{th}$ and there are four unimportant feature maps which are $9 \mathrm{th}$, $12 \mathrm{th}$, $16 \mathrm{th}$, $17 \mathrm{th}$. And it can be seen that the attention of each category is similar in most of feature maps. It may suggest that different types of signals shared channel features in the early stages of the network. As the depth of the network increases, the attention of feature channels are different and the advantage of the recalibration is less important. 

Interestingly, in the early stages of the network, important convolution kernels were few and shared by most categories. In other words, if SA is used in the early stages of the network, the network will calibrate feature for different modulation modes. In the Fig. \ref{fig_sim5} (b), different feature maps has higher weights and are similar for each modulation type. We can get a result, where the lower level of feature is more important and this suggests that the higher level of features are less important than previous blocks in providing recalibration to the network.

\section{Conclusion} 

In order to improve the recognition rate in AMR and reduce the complexity of the model, a new network structure called Fully Dense Neural Network (FDNN) is proposed in this paper. We design dense connections for residual blocks to ensure efficient feature extraction ,while encouraging feature reuse to reduce a large of trainable parameters in model. Experimental results on the RadioML2016.10a dataset shows that our approach can extract efficient feature and reduce the overfitting. Meanwhile, the experiment suggest that dense connections is a very promising method in AMR, which can not only greatly reduce training parameters but also improve recognition performance. Furthermore, we analyze the role of attention and the importance of feature layers in automatic modulation recognition. Experimental results shows that the attention mechanism can recalibrate the network to enhance features. Finally, we conclude that in IQ data for AMR, the attention component is effective, and that residual blocks and feature reuse have advantages of extracting effective features and reducing model complexity.




%





\ifCLASSOPTIONcaptionsoff
  \newpage
\fi

\bibliographystyle{ieeetr}
\bibliography{FDNN}

\end{document}